# Hybrid entanglement caused by depletion of pump on output of spontaneous parametric down converter

Sergey A. Podoshvedov

*Department of computer modeling and nanotechnology, Institute of natural and exact sciences, South Ural State University, Lenin Av. 76, Chelyabinsk, Russia*
e-mail: sapodo68@gmail.com

We consider novel method for implementation of hybrid entanglement between microscopic and macroscopic states on output of spontaneous parametric down converter through the depletion of the pump wave. The generated signal, idler and pumping fields show weak entanglement. We characterize it and show the limits of applicability of the depleted pump regime. Testing method of the hybrid entanglement is proposed. The hybrid entanglement is tested by conditional generation of maximally entangled states of two qubits and qutrits regardless of its smallness. Possibilities to enhance the hybrid entanglement and, as consequence, success probabilities of the conditional generation are considered. The optical scheme is realizable in practice and can do without photon number resolving detection.

## 1. Introduction

Quantum entanglement is key ingredient for fundamental tests of quantum mechanics [1-3] and implementations of quantum information processing [4-7]. Entanglement between two inherently different parts is often referred to hybrid. The subsystems of composite system can differ in their nature (an electromagnetic field and an atomic system), their size (microscopic and macroscopic systems) or in the way they can be described (discrete- and continuous variable description of physical systems). Schrödinger's famous cat paradox [8], where classical object (cat) is entangled with quantum (quantum particle) can serve example of the hybrid entanglement between the microscopic quantum and macroscopic classical physical systems. Also, the implementation of the hybrid states can be of enough interest for practical purposes. The standard idea to implement entangling gates which along with the one-qubit transformations are the basic elements of quantum computing is based on the teleportation protocol [6] and Bell-state measurement [7,9]. But the problem in implementation the two-qubit gates with linear optics elements and photodetectors is that the success probability of the Bell-state measurement does not exceed 0.5 [10-12]. As result, the success probability of the controlled operations like to controlled $-X$ operation performed by simultaneous teleportation of two arbitrary qubits through entangled quantum channel is limited to 0.25 [13-15]. Hybrid entangled states may be important tool for quantum technologies [16]. Now, the study of the hybrid entanglement between macroscopic and microscopic states has become the subject of intense as both experimental [17,18] and theoretical researches [19]. Recently, some implementations of the hybrid entanglement between a coherent qubit (superposition of coherent states) and microscopic qubit of vacuum and single photon [20] and a single photon in polarization basis [21] were demonstrated.

Here, we develop novel way to implement quantum entanglement between microscopic and macroscopic states on the output from spontaneous parametric down converter (SPDC). The entanglement between the states is caused by inverse transformation of the pump photon into photons in the signal and idler modes. Consideration of the hybrid entanglement generation requires SPDC with depleted pumping wave [22,23]. The generated hybrid entanglement is turned out to be rather weak and depends on the coupling parameter which along with squeezing parameter completely determines the output state of the SPDC. If we neglect the contribution of the terms which are proportional to the coupling parameter, the two-mode squeezed state [24] is observed at the output of the nonlinear crystal. In spite of the smallness of the generated hybrid entanglement, it can be tested. The hybrid entanglement can be used to conditionally prepare maximally entangled states of qubits, qutrits and qudits. A detailed description of SPDC in depleted pump regime is presented in section 2. It is shown generation of the hybrid entanglement on the output of the SPDC. In section 3, we describe a method to test the hybrid entanglement through the conditional generation of maximally entangled states. Section 4 is devoted to the discussion of the main moments of the entanglement, its comparison with others. Different strategies for increasing the hybrid entanglement are discussed. Additional auxiliary mathematical is presented in Appendixes A and B.

## 2. Spontaneous parametric down conversion with depleted pump and generation of hybrid entanglement

Consider the three-mode model of the interaction of the signal, idler and pumping light fields in crystal with second-order nonlinearity $\chi^{(2)}$. In interaction representation, the model, when we neglect influence of decoherence, is described by the Hamiltonian [22,23,25]

$$H = i\hbar r \left( a_1^+ a_2^+ a_p - a_p^+ a_1 a_2 \right), \tag{1}$$

where $a$, $a^+$ are the bosonic annihilation and creation operators for the number states $|n\rangle$ $\left( a|n\rangle = \sqrt{n}|n-1\rangle, a^+|n\rangle = \sqrt{n+1}|n+1\rangle \right)$, subscripts $1, 2, p$ account for generated signal, idler and pumping field modes, respectively, $r$ is a coupling parameter connecting the modes and $\hbar$ is Planck constant. A nonlinear crystal with nonlinearity $\chi^{(2)}$ is used to split pumping photon into pairs in accordance with the laws of conservation of energy and momentum. For a continuous-wave pump at frequency $\omega_p$, the signal and idler photons are generated with frequencies satisfying the energy-conservation relation $\omega_p = \omega_1 + \omega_2$. The momentum conservation or phase matching condition $\vec{k}_p = \vec{k}_1 + \vec{k}_2$, where $\vec{k}_p$, $\vec{k}_1$, $\vec{k}_2$ are the wave vectors of the corresponding light fields, is valid inside the crystal. The model (1) can be used for type-I phase matching, when a pump beam generates the signal and idler photons of the same polarization. In particular, we can consider near-deterministic SPDC, when phase matching is achieved for photons with frequencies approximately half of the pump frequency. Whereas degenerate SPDS can also occur in the bulk crystal it can be excluded through frequency filtering at the output. The same model can be applicable to type-II phase-matched degenerate (or the near-degenerate) process, when a pumping photon generates photons of the same frequency (or approximately the same frequency) of half of the pump with mutually perpendicular polarizations. So, at certain conditions, the generated down-converted photons diverge from each other and from the direction of the pump mode in bulk structure. To satisfy the phase matching conditions the photons are emitted along cones. If the cones are intersected along two directions, then the two-photon state with uncertain polarization or the polarization entangled state can be generated [26]. Whereas down conversion can also occur in wide diapason of parameters, the other modes can also be removed by frequency and spatial filtering that makes correct the three-mode Hamiltonian description (1).

Usually, the Hamiltonian (1) is considered in undepleted pump regime which means that the input pump-beam profile amplitude remains constant during the interaction. This implies that only the pump photons are converted into signal and idler ones. Conversion of the signal and idler photons back into the pump photon is not considered as it is believed that contribution of the transformation to light fields is too negligible. The consideration is quite reasonable because the pumping field is a macroscopic state (visible by the eye [17,18]) while the generated states are microscopic ones. Use of the Hamiltonian (1) makes it possible to take into account not only the impact of the macroscopic field on generated microscopic states but reverse influence of the signal and idler fields on the pump. Consideration of such influence allows us to consider the possibility of generating a hybrid entanglement between macroscopic and microscopic light states. Low input intensity in the pumping mode generates biphoton on output being a pair of photons correlated in moments of their birth [27]. High degree of correlation between the photon number states in each mode is observed in two-mode squeezed vacuum (TMSV) [25]

$$|\Psi\rangle_{12} = S_{12}(r)|00\rangle_{12} = (1/\cosh r)\sum_{n=0}^{\infty}(\tanh r)^n |n\rangle_1 |n\rangle_2, \tag{2}$$

where the two-mode operator $S_{12}(r)$ is given by

$$S_{12}(r) = \exp(r(a_1^+ a_2^+ - a_1 a_2)) \tag{3}$$

with $r$ being the squeezing parameter and eigenstates $|n\rangle$ of Hamiltonian of quantum harmonic oscillator are orthogonal $\langle n|m\rangle = \delta_{nm}$, where $\delta_{nm} = 1$, if $n = m$, otherwise $\delta_{nm} = 0$. TMSV is generated due to interaction of light fields with $\chi^{(2)}$ nonlinearity, contains multiple photon pairs and can be result in largely entangled state. The entanglement can be detected with homodyne detection by means of the Duan-Simon criterion [28,29]. TMSV (2) can be considered with moderate values of the squeezing parameter $r \leq 1$. A further increase of the coupling parameter can lead to that the approach (2) is no longer adequate to describe the output state as the signal and idler modes may become entangled with the state in the pumping mode.

Initial state to the Hamiltonian (1) is

$$|\Psi_{In}\rangle_{12p} = |00\rangle_{12}|0,\alpha\rangle_p, \tag{4}$$

that means the signal and idler modes are in vacuum and the pumping mode in coherent state $|0,\alpha\rangle$ is applied with parameter $\alpha$ being an amplitude of the coherent state. Here, we have used the same notations as in [30] for designation of the displaced number states with discrete number $n$ and continuous parameter (size) $\alpha$ which shows on which amplitude the number state is displaced on the phase plane. If we take into account multiple photon pairs generation and depletion of the pump, then we can finally write output entangled state as [23]

$$|\Psi_{Out}\rangle_{12p} = \sum_{n=0}^{\infty}(\alpha\eta)^n |nn\rangle_{12}|\Phi_n^{(00)}\rangle_p, \tag{5}$$

where the states in the pumping modes are labeled by the subscript $n$ and superscript 00 that implies the states are generated when the signal and idler modes are initially in vacuum state. Coupling parameter $\eta$ is $rc/L$ ($\eta = rc/L$), where $L$ is a crystal length. Bright unnormalized states in the pump mode are the following

$$|\Phi_n^{(00)}\rangle = \exp(-|\alpha|^2/2)\sum_{l=0}^{\infty}\frac{\alpha^{l+n}}{\sqrt{(l+n)!}} f_{2(l+n),n+1}^{(00)} |l\rangle. \tag{6}$$

The states $|\Phi_n^{(00)}\rangle$ are characterized by the matrix elements $f_{2l,k}^{(00)}$ being real quantities that obey system of $k+1$ linear differential equations [22,23]

$$\frac{df_{2l,k}^{(00)}}{d\tau} = \eta\left((k-1)\sqrt{l-k+2}f_{2l,k-1}^{(00)} - k\sqrt{l-k+1}f_{2l,k+1}^{(00)}\right), \tag{7}$$

where dimensionless time $\tau = tc/L$ is changed from 0 to 1. Here, the first argument $l$ of the elements $f_{2l,k}^{(00)}$ can take values from 0 up to $\infty$ and the second argument $k$ is varied in diapason from 1 to $l+1$. Input column-vector of the matrix elements satisfies the condition

$$| f_{2l,1}^{(00)}(0) \quad f_{2l,2}^{(00)}(0) \quad f_{2l,3}^{(00)}(0) \quad ... \quad f_{2l,l+1}^{(00)}(0) |^T = | 1 \quad 0 \quad 0 \quad ... \quad 0 |^T, \quad (8)$$

where symbol $T$ means matrix transposition, that corresponds to initial coherent state (4). Normalization condition $\langle \Psi_{Out} | \Psi_{Out} \rangle_{12p} = 1$ for the output state (5) is kept and the following equality takes place

$$\exp(-|\alpha|^2) \sum_{n=0}^{\infty} |\alpha\eta|^{2n} \sum_{l=0}^{\infty} \frac{|\alpha^{l+n}|^2}{(l+n)!} f_{2(l+n),n+1}^{(00)2}(\tau) = 1. \quad (9)$$

The states (6) represent infinite series of the number state with amplitudes that may resemble the coherent state. We can also consider it as decomposition over displaced number states $|n,\alpha\rangle$ with the displacement amplitude $\alpha$ ($\alpha$ – representation) [31]. Summing up results of Appendix A (A2), we can rewrite unnnormalized states (6) in the pumping modes as

$$\left| \Phi_n^{(00)} \right\rangle = \sum_{m=0}^{\infty} g_{nm} |m,\alpha\rangle, \quad (10)$$

for completeness of the set of the displaced number states (10) for any value of $\alpha$ [30,31]. The normalization condition (9) is kept for the amplitudes $g_{nm}$

$$\sum_{n=0}^{\infty} \sum_{m=0}^{\infty} |\alpha\eta|^{2n} |g_{nm}|^2 = 1. \quad (11)$$

The matrix elements $g_{nm}$ (A10-A19) are the wave amplitudes of the state (10). The matrix elements of the state with arbitrary number $n$ (10) have a common property. The amplitudes $g_{nm}$ for the state $\left| \Phi_n^{(00)} \right\rangle$ or the same matrix elements of $m$ row of the matrix (A3) are product of the parameter $\eta$ in power $m$ $(\eta^m)$ by some expression enclosed in parenthesis. The additional expressions are the same order and they are composed of the terms comprising squeezing parameter $\alpha\eta$, coupling parameter $\eta$ and some numerical expressions. We estimate relation between two neighboring matrix elements $g_{nm}$ and $g_{nm+1}$ regardless on number $n$ as

$$\frac{|g_{nm}|}{|g_{nm+1}|} \sim \frac{1}{\eta}. \quad (12)$$

It implies that the amplitude of the displaced number state $|n,\alpha\rangle$ in the superposition (10) is greater of the amplitude of the subsequent term $|n+1,\alpha\rangle$ in $1/\eta \gg 1$ times. As contribution of the displaced state with $m=0$ prevails over the contribution of the states with larger quantum number in the superposition (10), in the undepleted pump regime, it is possible to neglect all the matrix elements by taking them equal to zero $g_{nm}=0$ for all values of $m$ with the exception of $m=0$ $g_{n0} \neq 0$. Then, we present the state (5) as

$$\left| \Psi_{Out} \right\rangle_{12p} = N_1 \sum_{n=0}^{\infty} (\alpha\eta)^n g_{n0} |nn\rangle_{12} |0,\alpha\rangle_p, \quad (13)$$

where the normalization factor

$$N_1 = \left( \sum_{n=0}^{\infty} |(\alpha\eta)^n g_{n0}|^2 \right)^{-1/2}$$

is introduced. The state corresponds to the TMSV (2). We can also consider the case of SPDC with $\alpha\eta \ll 1$ to estimate the matrix elements $g_{n0}$ to be equal to one $(g_{n0} = 1)$ (Eqs. (A10,A14,A17)). It allows us to rewrite (13) as

$$|\Psi_{Out}\rangle_{12p} = N_2 \sum_{n=0}^{\infty} (\alpha\eta)^n |nn\rangle_{12} |0,\alpha\rangle_p , \qquad (14)$$

where $N_2 = \left(\sum_{n=0}^{\infty} |\alpha\eta|^{2n}\right)^{-1/2}$ is the corresponding normalization factor. The same expression for the output squeezed state follows from (2) in the case of $r \ll 1$. We note only that the classic amplitude of the undepleted pump is already included in the parameter $r$.

In general case, a set of macroscopic (bright) states (10) is not orthogonal $\langle \Phi_m^{(00)} | \Phi_n^{(00)} \rangle \neq 0$ with $m \neq n$ but nevertheless they are not identical to each other as every of the macroscopic states represents an infinite superposition of the displaced number states with different amplitudes. The output state (5) is entangled. Photons generated in the signal and idler modes are correlated with the state in the pumping mode. The state can be recognized hybrid as a state (10) can be considered as a macroscopic due to big value of displacement $\alpha$ and the states of the generated photons are microscopic. The state in the signal and idler modes is obtained by tracing over the displaced states in the pumping mode and becomes

$$\rho_{12} = Sp\left(|\Psi_{Out}\rangle_{12p\ 12p}\langle\Psi_{Out}|\right)_p = |00\rangle\langle 00| + \alpha\eta(J_{01}|00\rangle\langle 11| + J_{10}|11\rangle\langle 00|) + \\ (\alpha\eta)^2(J_{02}|00\rangle\langle 22| + J_{11}|11\rangle\langle 11| + J_{20}|22\rangle\langle 00|) + ... \qquad (15)$$

where summation is produced over the displaced states in the pumping mode, $J_{nm} = \sum_{k=0}^{\infty} g_{nk} g_{mk}^*$, $J_{nm} = J_{nm}^*$. The correlation matrix $J$ can be derived with help of matrix $G$ (A3). The state (15) is mixed as the hybrid entanglement on output of SPDC exists. The hybrid entanglement between the generated signal idler photons and state in the pumping mode is determined by the coupling parameter $\eta$. If we neglect all terms containing $\eta \ll 0$ in expressions for the amplitudes $g_{nm}$, the hybrid entanglement disappears. The state (5) as whole becomes separate in the case but the generated photons in signal and idler modes remain entangled and measure of the entanglement can be estimated [28,29]. The case corresponds to generation of two-mode squeezed states described by the expressions (2,13,14). Thus, output state (5) of the SPDC is determined by two parameters: the squeezing parameter $r = \alpha\eta$ and the coupling parameter $\eta$. Squeezing parameter $r = \alpha\eta$ is connected with correlation properties of TMSV and it accounts for how strong noise in one of the quadrature components of the light field can be squeezed. Parameter $\eta$ is responsible for existence of the matrix elements $g_{nm}$ ($m \neq 0$) in matrix (A3) and, as consequence, for generation of the hybrid entanglement between signal, idler and pumping modes. An increase of the coupling parameter $\eta$ leads to an increase of the hybrid entanglement.

**3. Test of the hybrid entanglement by conditional generation of maximally entangled states**

As rule, the coupling coefficient $\eta$ is much less of one $\eta \ll 1$ in real situation. Therefore, it is logical to neglect the hybrid entanglement (5) as it is not so easy to contrive the way to use it in practice. Another problem may connected with that how practically to identify the hybrid entanglement. Here, we propose approach directed on detection of the hybrid entanglement. Manifestation of the hybrid entanglement in the output state (5) can be traced through conditional generation of the maximally entangled states. The conditional generation of maximally entangled states can be considered as criterion of existence of the hybrid entanglement on output from the SPDC. Note only, we are going to use the following output unnormalized state

$$|\Psi_{Out}\rangle_{12p} = |00\rangle_{12}|\Phi_0^{(00)}\rangle_p + (\alpha\eta)|11\rangle_{12}|\Phi_1^{(00)}\rangle_p + (\alpha\eta)^2|22\rangle_{12}|\Phi_2^{(00)}\rangle_p, \qquad (16a)$$

instead of (5) for mathematical calculations that is reasonable in the case of $\alpha\eta < 1$. The calculation can be even done for more simplified output state consisting of two terms

$$|\Psi_{Out}\rangle_{12p} = |00\rangle_{12}|\Phi_0^{(00)}\rangle_p + (\alpha\eta)|11\rangle_{12}|\Phi_1^{(00)}\rangle_p, \qquad (16b)$$

in the case of $\alpha\eta << 1$. The state (16b) to the greatest degree could be recognized unbalanced hybrid provided that the states $|\Phi_0^{(00)}\rangle$ and $|\Phi_1^{(00)}\rangle$ are orthogonal. However, the weak hybrid entanglement is enough for generation of the maximally entangled states composed from qubits and qutrits. We use the form (16a) instead of (16b) to improve the accuracy of the results.

Consider the tensor product of two entangled states (5) $|\Psi_{Out}\rangle_{12p_1}$ and $|\Psi_{Out}\rangle_{34p_2}$ generated by two SPDC (Fig. 1(a)) with correlated states in the pumping modes. The state can be represented in series of increasing powers of the squeezing parameter $(\alpha\eta)^n$ with $n$ varying from $0$ up to $\infty$. The coefficients of this series are endless blocks involving both microscopic and macroscopic states that we are going to denote by $F_n^{(2)}$ leaving out ket vector notation $|\ \rangle$ for them. Finally, it can be rewritten as infinite superposition

$$|\Psi_{Out}\rangle_{12p_1}|\Psi_{Out}\rangle_{34p_2} = F_0^{(2)} + \alpha\eta F_1^{(2)} + (\alpha\eta)^2 F_2^{(2)} + (\alpha\eta)^3 F_3^{(2)} + \ldots, \qquad (17)$$

where first three blocks are given by

$$F_0^{(2)} = |0000\rangle_{1234}\left(\begin{array}{l} p_{00}^{(2)}|0,\alpha\rangle_{p_1}|0,\alpha\rangle_{p_2} + p_{01}^{(2)}|\varphi_+\rangle_{p_1p_2} + \\ p_{02}^{(2)}|\sigma_+\rangle_{p_1p_2} + p_{03}^{(2)}|1,\alpha\rangle_{p_1}|1,\alpha\rangle_{p_2} + \ldots\end{array}\right), \qquad (18a)$$

$$F_1^{(2)} = p_{10}^{(2)}|\Psi_+\rangle_{1234}|0,\alpha\rangle_{p_1}|0,\alpha\rangle_{p_2} + p_{11}^{(2)}|\Psi_+\rangle_{1234}|\varphi_+\rangle_{p_1p_2} + p_{12}^{(2)}|\Psi_-\rangle_{1234}|\varphi_-\rangle_{p_1p_2}$$
$$+ p_{13}^{(2)}|\Psi_+\rangle_{1234}|\sigma_+\rangle_{p_1p_2} + p_{14}^{(2)}|\Psi_-\rangle_{1234}|\sigma_-\rangle_{p_1p_2} + p_{15}^{(2)}|\Psi_+\rangle_{1234}|1,\alpha\rangle_{p_1}|1,\alpha\rangle_{p_2} + \ldots, \qquad (18b)$$

$$F_2^{(2)} = p_{20}^{(2)}|\Phi_+\rangle_{1234}|0,\alpha\rangle_{p_1}|0,\alpha\rangle_{p_2} + p_{21}^{(2)}|\Phi_+\rangle_{1234}|\varphi_+\rangle_{p_1p_2} + p_{22}^{(2)}|\Phi_-\rangle_{1234}|\varphi_-\rangle_{p_1p_2} +$$
$$p_{23}^{(2)}|\Phi_+\rangle_{1234}|\sigma_+\rangle_{p_1p_2} + p_{24}^{(2)}|\Phi_-\rangle_{1234}|\sigma_-\rangle_{p_1p_2} + p_{25}^{(2)}|\Phi_+\rangle_{1234}|1,\alpha\rangle_{p_1}|1,\alpha\rangle_{p_2} + \ldots \qquad (18c)$$
$$|1111\rangle_{1234}\left(p_{26}^{(2)}|0,\alpha\rangle_{p_1}|0,\alpha\rangle_{p_2} + p_{27}^{(2)}|\varphi_+\rangle_{p_1p_2} + p_{28}^{(2)}|\sigma_+\rangle_{p_1p_2} + p_{29}^{(2)}|1,\alpha\rangle_{p_1}|1,\alpha\rangle_{p_2} + \ldots\right)$$

Here superscript $(2)$ concerns tensor product of two states. The entangled microscopic states in the generated modes

$$|\Psi_\pm\rangle_{1234} = (|1100\rangle \pm |0011\rangle)_{1234}/\sqrt{2}, \qquad (19a)$$
$$|\Phi_\pm\rangle_{1234} = (|2200\rangle \pm |0022\rangle)_{1234}/\sqrt{2}, \qquad (19b)$$

and macroscopic states in the pumping modes

$$|\varphi_\pm\rangle_{p_1p_2} = (|0,\alpha\rangle|1,\alpha\rangle \pm |1,\alpha\rangle|0,\alpha\rangle)_{p_1p_2}/\sqrt{2}, \qquad (20a)$$
$$|\sigma_\pm\rangle_{p_1p_2} = (|0,\alpha\rangle|2,\alpha\rangle \pm |2,\alpha\rangle|0,\alpha\rangle)_{p_1p_2}/\sqrt{2}. \qquad (20b)$$

are introduced. Integer $n$ or the same power of the factor $\alpha\eta$ characterizes the block. The blocks with $n > 2$ are not presented in the expression (17) as their contributions may be negligible in the case of $\alpha\eta < 1$ as in (16a). Moreover, each block can be divided into sub-blocks, which are enclosed in parentheses in the block. For example, the block which is characterized by an integer $n=2$ (19c) involves two sub-blocks. One of them is in parentheses with state $|1111\rangle_{1234}$ in front of it inside the block. This form of the tensor product of two states (17) is obtained by unitary conversion from the basic states to entangled ones (19-20). Amplitudes of the members are converted in the same manner as basic states. The amplitudes are presented in (A20-A39). The amplitudes obey the normalization condition

$$\sum_{n=0}^{\infty}\sum_{m=0}^{\infty}|\alpha\eta|^{2n}|p_{nm}^{(2)}|^2 = 1. \tag{21}$$

Consider the tensor product of three identical output states (5) $|\Psi_{Out}\rangle_{12p_1}$, $|\Psi_{Out}\rangle_{34p_2}$ and $|\Psi_{Out}\rangle_{56p_3}$ with correlated pumping in all modes (Fig. 1(b)). Following the same calculation algorithm as in (17), we can rewrite the state in block form

$$|\Psi_{Out}\rangle_{12p_1}|\Psi_{Out}\rangle_{34p_2}|\Psi_{Out}\rangle_{56p_3} = F_0^{(3)} + \alpha\eta F_1^{(3)} + (\alpha\eta)^2 F_2^{(3)} + (\alpha\eta)^3 F_3^{(3)} + ..., \tag{22}$$

where blocks with superscript $(3)$ indicate on triple tensor product. They are given by

$$F_0^{(3)} = |000000\rangle_{123456}\begin{pmatrix} p_{00}^{(3)}|0,\alpha\rangle_{p_1}|0,\alpha\rangle_{p_2}|0,\alpha\rangle_{p_3} + p_{01}^{(3)}|\varphi_0\rangle_{p_1p_2p_3} + \\ p_{02}^{(3)}|\sigma_0\rangle_{p_1p_2p_3} + p_{03}^{(3)}|\xi_0\rangle_{p_1p_2p_3} + ... \end{pmatrix}, \tag{23a}$$

$$\begin{aligned}F_1^{(3)} &= p_{10}^{(3)}|\Psi_0\rangle_{123456}|0,\alpha\rangle_{p_1}|0,\alpha\rangle_{p_2}|0,\alpha\rangle_{p_3} + p_{11}^{(3)}|\Psi_0\rangle_{123456}|\varphi_0\rangle_{p_1p_2p_3} + \\ &\quad p_{12}^{(3)}|\Psi_\phi\rangle_{123456}|\varphi_{-\phi}\rangle_{p_1p_2p_3} + p_{12}^{(3)}|\Psi_{2\phi}\rangle_{123456}|\varphi_{-2\phi}\rangle_{p_1p_2p_3} + \\ &\quad p_{13}^{(3)}|\Psi_0\rangle_{123456}|\sigma_0\rangle_{p_1p_2p_3} + p_{14}^{(3)}|\Psi_\phi\rangle_{123456}|\sigma_{-\phi}\rangle_{p_1p_2p_3} + p_{14}^{(3)}|\Psi_{2\phi}\rangle_{123456}|\sigma_{-2\phi}\rangle_{p_1p_2p_3} + \\ &\quad p_{15}^{(3)}|\Psi_0\rangle_{123456}|\xi_0\rangle_{p_1p_2p_3} + p_{16}^{(3)}|\Psi_\phi\rangle_{123456}|\xi_{-\phi}\rangle_{p_1p_2p_3} + p_{17}^{(3)}|\Psi_{2\phi}\rangle_{123456}|\xi_{-2\phi}\rangle_{p_1p_2p_3} + ...\end{aligned} \tag{23b}$$

$$\begin{aligned}F_2^{(3)} &= p_{20}^{(3)}|\Phi_0\rangle_{123456}|0,\alpha\rangle_{p_1}|0,\alpha\rangle_{p_2}|0,\alpha\rangle_{p_3} + p_{21}^{(3)}|\Phi_0\rangle_{123456}|\varphi_0\rangle_{p_1p_2p_3} + \\ &\quad p_{22}^{(3)}|\Phi_\phi\rangle_{123456}|\varphi_{-\phi}\rangle_{p_1p_2p_3} + p_{22}^{(3)}|\Phi_{2\phi}\rangle_{123456}|\varphi_{-2\phi}\rangle_{p_1p_2p_3} + \\ &\quad p_{23}^{(3)}|\Phi_0\rangle_{123456}|\sigma_0\rangle_{p_1p_2p_3} + p_{24}^{(3)}|\Phi_\phi\rangle_{123456}|\sigma_{-\phi}\rangle_{p_1p_2p_3} + p_{24}^{(3)}|\Phi_{2\phi}\rangle_{123456}|\sigma_{-2\phi}\rangle_{p_1p_2p_3} + \\ &\quad p_{25}^{(3)}|\Phi_0\rangle_{123456}|\xi_0\rangle_{p_1p_2p_3} + p_{26}^{(3)}|\Phi_\phi\rangle_{123456}|\xi_{-\phi}\rangle_{p_1p_2p_3} + p_{27}^{(3)}|\Phi_{2\phi}\rangle_{123456}|\xi_{-2\phi}\rangle_{p_1p_2p_3} + \\ &\quad p_{28}^{(3)}|\Delta_0\rangle_{123456}|0,\alpha\rangle_{p_1}|0,\alpha\rangle_{p_2}|0,\alpha\rangle_{p_3} + p_{29}^{(3)}|\Delta_0\rangle_{123456}|\varphi_0\rangle_{p_1p_2p_3} + \\ &\quad p_{210}^{(3)}|\Delta_\phi\rangle_{123456}|\varphi_{-\phi}\rangle_{p_1p_2p_3} + p_{211}^{(3)}|\Delta_{2\phi}\rangle_{123456}|\varphi_{-2\phi}\rangle_{p_1p_2p_3} + p_{212}^{(3)}|\Delta_0\rangle_{123456}|\sigma_0\rangle_{p_1p_2p_3} + \\ &\quad p_{213}^{(3)}|\Delta_\phi\rangle_{123456}|\sigma_{-\phi}\rangle_{p_1p_2p_3} + p_{214}^{(3)}|\Delta_{2\phi}\rangle_{123456}|\sigma_{-2\phi}\rangle_{p_1p_2p_3} + p_{215}^{(3)}|\Delta_0\rangle_{123456}|\xi_0\rangle_{p_1p_2p_3} + \\ &\quad p_{216}^{(3)}|\Delta_\phi\rangle_{123456}|\xi_{-\phi}\rangle_{p_1p_2p_3} + p_{216}^{(3)}|\Delta_{2\phi}\rangle_{123456}|\xi_{-2\phi}\rangle_{p_1p_2p_3} + ...\end{aligned} \tag{23c}$$

where triples of entangled microscopic states of the generated photons are introduced

$$|\Psi_0\rangle_{123456} = (|110000\rangle + |001100\rangle + |000011\rangle)_{123456}/\sqrt{3}, \tag{24a}$$

$$|\Psi_\phi\rangle_{123456} = (|110000\rangle + \exp(i\phi)|001100\rangle + \exp(i2\phi)|000011\rangle)_{123456}/\sqrt{3}, \tag{24b}$$

$$|\Psi_{2\phi}\rangle_{123456} = (|110000\rangle + \exp(i2\phi)|001100\rangle + \exp(i4\phi)|000011\rangle)_{123456}/\sqrt{3}, \tag{24c}$$

$$|\Phi_0\rangle_{123456} = (|220000\rangle + |002200\rangle + |000022\rangle)_{123456}/\sqrt{3}, \tag{25a}$$

$$|\Phi_\phi\rangle_{123456} = (|220000\rangle + \exp(i\phi)|002200\rangle + \exp(i2\phi)|000022\rangle)_{123456}/\sqrt{3}, \tag{25b}$$

$$|\Phi_{2\phi}\rangle_{123456} = (|220000\rangle + \exp(i2\phi)|002200\rangle + \exp(i4\phi)|000022\rangle)_{123456}/\sqrt{3}, \quad (25c)$$

$$|\Delta_0\rangle_{123456} = (|111100\rangle + |001111\rangle + |110011\rangle)_{123456}/\sqrt{3}, \quad (26a)$$

$$|\Delta_\phi\rangle_{123456} = (|111100\rangle + \exp(i\phi)|001111\rangle + \exp(i2\phi)|110011\rangle)_{123456}/\sqrt{3}, \quad (26b)$$

$$|\Delta_{2\phi}\rangle_{123456} = (|111100\rangle + \exp(i2\phi)|001111\rangle + \exp(i4\phi)|110011\rangle)_{123456}/\sqrt{3}, \quad (26c)$$

where phase shift is $\phi = 2\pi/3$. Macroscopic three-partite entangled states in the pumping mode are given by

$$|\varphi_0\rangle_{123} = (|1,\alpha\rangle|0,\alpha\rangle|0,\alpha\rangle + |0,\alpha\rangle|1,\alpha\rangle|0,\alpha\rangle + |0,\alpha\rangle|0,\alpha\rangle|1,\alpha\rangle)_{123456}/\sqrt{3}, \quad (27a)$$

$$|\varphi_{-\phi}\rangle_{123} = \begin{pmatrix} |1,\alpha\rangle|0,\alpha\rangle|0,\alpha\rangle + \exp(-i\phi)|0,\alpha\rangle|1,\alpha\rangle|0,\alpha\rangle + \\ \exp(-i2\phi)|0,\alpha\rangle|0,\alpha\rangle|1,\alpha\rangle \end{pmatrix}_{123456}/\sqrt{3}, \quad (27b)$$

$$|\varphi_{-2\phi}\rangle_{123} = \begin{pmatrix} |1,\alpha\rangle|0,\alpha\rangle|0,\alpha\rangle + \exp(-i2\phi)|0,\alpha\rangle|1,\alpha\rangle|0,\alpha\rangle + \\ \exp(-i4\phi)|0,\alpha\rangle|0,\alpha\rangle|1,\alpha\rangle \end{pmatrix}_{123456}/\sqrt{3}, \quad (27c)$$

$$|\sigma_0\rangle_{123} = (|2,\alpha\rangle|0,\alpha\rangle|0,\alpha\rangle + |0,\alpha\rangle|2,\alpha\rangle|0,\alpha\rangle + |0,\alpha\rangle|0,\alpha\rangle|2,\alpha\rangle)_{123456}/\sqrt{3}, \quad (28a)$$

$$|\sigma_{-\phi}\rangle_{123} = \begin{pmatrix} |2,\alpha\rangle|0,\alpha\rangle|0,\alpha\rangle + \exp(-i\phi)|0,\alpha\rangle|2,\alpha\rangle|0,\alpha\rangle + \\ \exp(-i2\phi)|0,\alpha\rangle|0,\alpha\rangle|2,\alpha\rangle \end{pmatrix}_{123456}/\sqrt{3}, \quad (28b)$$

$$|\sigma_{-2\phi}\rangle_{123} = \begin{pmatrix} |2,\alpha\rangle|0,\alpha\rangle|0,\alpha\rangle + \exp(-i2\phi)|0,\alpha\rangle|2,\alpha\rangle|0,\alpha\rangle + \\ \exp(-i4\phi)|0,\alpha\rangle|0,\alpha\rangle|2,\alpha\rangle \end{pmatrix}_{123456}/\sqrt{3}, \quad (28c)$$

$$|\zeta_0\rangle_{123} = (|1,\alpha\rangle|1,\alpha\rangle|0,\alpha\rangle + |0,\alpha\rangle|1,\alpha\rangle|1,\alpha\rangle + |1,\alpha\rangle|0,\alpha\rangle|1,\alpha\rangle)_{123456}/\sqrt{3}, \quad (29a)$$

$$|\zeta_{-\phi}\rangle_{123} = \begin{pmatrix} |1,\alpha\rangle|1,\alpha\rangle|0,\alpha\rangle + \exp(-i\phi)|0,\alpha\rangle|1,\alpha\rangle|1,\alpha\rangle + \\ \exp(-i2\phi)|1,\alpha\rangle|0,\alpha\rangle|1,\alpha\rangle \end{pmatrix}_{123456}/\sqrt{3}, \quad (29b)$$

$$|\zeta_{-2\phi}\rangle_{123} = \begin{pmatrix} |1,\alpha\rangle|1,\alpha\rangle|0,\alpha\rangle + \exp(-i2\phi)|0,\alpha\rangle|1,\alpha\rangle|1,\alpha\rangle + \\ \exp(-i4\phi)|1,\alpha\rangle|0,\alpha\rangle|1,\alpha\rangle \end{pmatrix}_{123456}/\sqrt{3}. \quad (29c)$$

Amplitudes of the members in the blocks (23) are presented in (A40-A68). They satisfy normalization condition

$$\sum_{n=0}^{\infty}\sum_{m=0}^{\infty}|\alpha\eta|^{2n}|p_{nm}^{(3)}|^2 = 1. \quad (30)$$

We can construct the tensor product with bigger number $(n > 3)$ of the output states by the same manner as shown above.

Final states composed of two (17) or three (22) output states of the SPDC are the basis for the generation of the conditional states. Measurement-based linear optical quantum generation of new states looks attractive. If a measurement is performed on a portion of a composite system, the output state of unmeasured part of the correlated system strongly depends on the result of the measurement. It provides an alternative mechanism to effectively achieve nonlinear effect comparable with that of nonlinear media. Correlation between pumping and generated modes takes place due to coupling parameter $\eta$. Consider if on example of the state (17). We use the following unitary transformation in the pumping modes in Fig. 1(a)

$$BS = \frac{1}{\sqrt{2}}\begin{vmatrix} 1 & 1 \\ -1 & 1 \end{vmatrix} \quad (31)$$

realized by the balanced beam splitter followed by measurement in first pumping mode. Outcomes of the unitary transformation are presented in (B1-B6) Note only that coherent part of the states with amplitude $\sqrt{2}\alpha$ arises in second pumping mode while the states in the first pumping mode get rid of it. Measurement can be done by avalanche photodiodes (APD) not using photon number resolving detection (PNRD) able to distinguish outcomes from different number states. APD possesses high quantum efficiency but can only discriminate the presence of radiation from the vacuum. It can be used to reconstruct the photon statistics but cannot be used as photon counters.

We are interested only in observing click in the first pumping mode $p_1$ as shown in Fig. 1(a). Single photon detection enables to register a photon information about which of the two pumping modes it arrived is erased despite the fact that one of these events have definitely occurred. Measurement by APD gives birth to conditional state being mixed state described by the density matrix

$$\rho^{(2)} = \frac{\left(P_1^{(2)}\left|\Psi_1^{(2)}\right\rangle\left\langle\Psi_1^{(2)}\right| + P_2^{(2)}\left|\Psi_2^{(2)}\right\rangle\left\langle\Psi_2^{(2)}\right|\right)}{P^{(2)}}, \tag{32}$$

where the conditional states (superscript $2$ in used to label them) are the following

$$\left|\Psi_1^{(2)}\right\rangle = \frac{p_{12}^{(2)}\left|\Psi_-\right\rangle_{1234} + (\alpha\eta)p_{22}^{(2)}\left|\Phi_-\right\rangle_{1234}}{\sqrt{N_1^{(2)}}}, \tag{33}$$

$$\left|\Psi_2^{(2)}\right\rangle = \frac{1}{\sqrt{N_2^{(2)}}}\left(\begin{array}{l}\left(p_{02}^{(2)} - p_{03}^{(2)}\right)\left|0000\right\rangle_{1234} + (\alpha\eta)\left(p_{13}^{(2)} - p_{15}^{(2)}\right)\left|\Psi_+\right\rangle_{1234} + \\ (\alpha\eta)^2\left(p_{23}^{(2)} - p_{25}^{(2)}\right)\left|\Phi_+\right\rangle_{1234} + (\alpha\eta)^2\left(p_{28}^{(2)} - p_{29}^{(2)}\right)\left|1111\right\rangle_{1234}\end{array}\right), \tag{34}$$

with probabilities to generate the states

$$P_1^{(2)} = |\alpha\eta|^2\left(\left|p_{12}^{(2)}\right|^2 + |\alpha\eta|^2\left|p_{22}^{(2)}\right|^2\right), \tag{35a}$$

$$P_2^{(2)} = \frac{1}{2}\left(\left|p_{02}^{(2)} - p_{03}^{(2)}\right|^2 + |\alpha\eta|^2\left|p_{13}^{(2)} - p_{15}^{(2)}\right|^2 + |\alpha\eta|^4\left|p_{23}^{(2)} - p_{25}^{(2)}\right|^2 + |\alpha\eta|^4\left|p_{28}^{(2)} - p_{29}^{(2)}\right|^2\right), \tag{35b}$$

$$P^{(2)} = P_1^{(2)} + P_2^{(2)}. \tag{35c}$$

Here, the quantities $N_1^{(2)}$ and $N_2^{(2)}$ are normalization factors of the states (33,34), respectively. The states (33), (34) are orthogonal to each other. We can estimate the probabilities (35a), (35b) relative to each other by making use of definition of the elements $p_{ij}^{(2)}$ (A22,A23,A26,A27,A29,A32,A33,A35,A38,A39) and $g_{ij}$ (A10-A19). Analysis shows $P_1^{(2)}/P_2^{(2)} \sim 1/\eta^2 \gg 1$ that guarantees the contribution of the state (34) in (32) is negligible. Thus, registration of one click in the pumping mode enables to generate the state (33) with fidelity almost equal to one $F = Sp\left(\rho^{(2)}\left|\Psi_1^{(2)}\right\rangle\left\langle\Psi_1^{(2)}\right|\right) \cong 1$. The form of the state (33) is determined by relation of their amplitudes $p_{12}^{(2)}/(\alpha\eta p_{22}^{(2)})$. The conditional state becomes maximally entangled state of two photons (two qubits) in four modes

$$\left|\Psi_1^{(2)}\right\rangle = \left|\Psi_-\right\rangle_{1234} \tag{36}$$

with nearly unit fidelity $F^{(2)} = \left|_{1234}\left\langle\Psi_1^{(2)}\middle|\Psi_-\right\rangle_{1234}\right|^2 \cong 1$ in the case of $\alpha\eta \ll 1$. The four-mode maximally entangled state (36) can be converted into two-mode maximally polarization entangled state with help of polarization beam splitter. Note only that the conditional state (33) can also include members with more number of correlated photons if we extend the analysis to take into account the blocks with $n > 2$ in (17) in the case of $\alpha\eta \sim 1$. It is worth noting that the generation of the maximally entangled state (36) is possible due to the hybrid

entanglement between the microscopic and macroscopic states (5). Consideration of this generation is impossible in the case of neglect by the hybrid entanglement (Eqs. (13,14)).

The same analysis is applicable to the state (23) composed of three output states from SPDC (Fig. 1(b)). The three pumping modes undergo unitary transformation described by the matrix

$$U(\phi) = \frac{1}{\sqrt{3}} \begin{vmatrix} 1 & 1 & 1 \\ 1 & \exp(i\phi) & \exp(i2\phi) \\ 1 & \exp(i2\phi) & \exp(i4\phi) \end{vmatrix}, \quad (37)$$

before the measurement. The unitary transformation can be constructed from the beam splitters and phase shifters [32] and it is example of discrete Fourier transform [9]. The unitary operation (37) transforms states (27-29) as it is presented in Appendix B (B7-B16). Two APD are placed in second and third pumping modes to probabilistically generate desired states. The conditional states are generated when click is registered in either pumping modes. The conditional states are mixed. But contribution of the terms forming the mixed state is incomparable with each other. This allows us not to take in account the terms with much less probabilities as it is done in consideration of generation of entangled state (32). Finally, registration of click in second pumping mode generates a pure state

$$\left|\Psi_1^{(3)}\right\rangle = \frac{p_{12}^{(3)}\left|\Psi_\phi\right\rangle_{123456} + \alpha\eta\left(p_{22}^{(3)}\left|\Phi_\phi\right\rangle_{123456} + p_{210}^{(3)}\left|\Delta_\phi\right\rangle_{123456}\right)}{\sqrt{N_1^{(3)}}}, \quad (38)$$

while another conditional state is produced provided that only APD in third mode fixed the click

$$\left|\Psi_2^{(3)}\right\rangle = \frac{p_{12}^{(3)}\left|\Psi_{2\phi}\right\rangle_{123456} + \alpha\eta\left(p_{22}^{(3)}\left|\Phi_{2\phi}\right\rangle_{123456} + p_{211}^{(3)}\left|\Delta_{2\phi}\right\rangle_{123456}\right)}{\sqrt{N_2^{(3)}}}, \quad (39)$$

where $N_1^{(3)}$, $N_2^{(3)}$ are the normalization factors of the states. Contribution of the states $p_{22}^{(3)}\left|\Phi_\phi\right\rangle_{123456} + p_{210}^{(3)}\left|\Delta_\phi\right\rangle_{123456}$ and $p_{22}^{(3)}\left|\Phi_{2\phi}\right\rangle_{123456} + p_{211}^{(3)}\left|\Delta_{2\phi}\right\rangle_{123456}$ in (38,39) is defined by the squeezed factor $\alpha\eta$. If the squeezing parameter is much less of one $\alpha\eta \ll 1$, then we can count the conditional states (38,39) become

$$\left|\Psi_1^{(3)}\right\rangle = \left|\Psi_\phi\right\rangle_{123456}, \quad (40)$$

$$\left|\Psi_2^{(3)}\right\rangle = \left|\Psi_{2\phi}\right\rangle_{123456}, \quad (41)$$

with almost ideal unit fidelity $F_1^{(3)} = \left|{}_{123456}\left\langle\Psi_1^{(3)}\middle|\Psi_\phi\right\rangle_{123456}\right|^2 \cong 1$ and $F_2^{(3)} = \left|{}_{123456}\left\langle\Psi_2^{(3)}\middle|\Psi_{2\phi}\right\rangle_{123456}\right|^2 \cong 1$. The state (24b) and (24c) are entangled ones of two qutrits being the superposition state of single photon simultaneously located in three modes. The entangled states of two qutrits (40,41) occupy six modes.

## 4. Conclusion

We considered a principal possibility to generate hybrid entanglement (Eq. (5)) between microscopic and macroscopic states on output of the SPDC. The reason, for which the hybrid entanglement on output from SPDC is generated, is the depletion of the pump wave. As a rule, the depletion of the pump wave is not taken into account when considering the generation of entangled states. Study is usually limited to consideration either of maximally entangled states (biphotons states) [26,27] or the generation of squeezed states [24,25]. Reconversion of generated photon into the pump photons is taken into consideration in depleted pump regime. This reverse conversion of photons is responsible for the change of state in the pumping mode and ultimately leads to the generation of hybrid entanglement between microscopic and macroscopic states. Analysis shows that the hybrid entanglement is determined by the coupling parameter $\eta$ (along with well-known squeezing parameter $r = \alpha\eta$) and it is relatively weak due to the fact that $\eta \ll 1$. We have shown that the generated three-mode state is converted into two-mode entangled state (two-mode squeezed vacuum) if we neglect the contribution of the terms that are proportional to $\eta$ (Eqs. (13,14)). Finally, we have proved that despite the generated hybrid entanglement is weak, however it exists.

We also proposed principal possibility to test the hybrid entanglement. The hybrid entanglement can be responsible for conditional preparation of maximally entangled states. We considered a possibility to conditionally generate entangled state of two qubits (two single photons being a superposition of two separate modes) and entangled state of two qutrits existing as a superposition of three separate (orthogonal quantum states) modes. Success probabilities to implement the states are sufficiently low

$$P_2 = \left|p_{12}^{(2)}\right|^2 \sim \eta^2, \tag{42a}$$

$$P_3 = 2\left|p_{12}^{(3)}\right|^2 \sim 2\eta^2, \tag{42b}$$

where $P_2$ and $P_3$ are the probabilities to conditionally generate maximally entangled states of two qubits and qutrits, respectively, due to smallness of the generated hybrid entanglement. An additional factor 2 in (42b) arises as the two states $\left|\Psi_\phi\right\rangle_{123456}$ and $\left|\Psi_{2\phi}\right\rangle_{123456}$ (Eqs. (40,41)) can be generated and they can be converted into each other by methods of linear optics [32]. This method can be extended to generate entangled states of two qudits that may take any of $d$ base states ($2d$ separate modes). It can be done by increasing SPDC output states and using discrete Fourier transform [9,32] with subsequent measurement of single photon in the corresponding modes. Then, the success probability of generation of the state is proportional to $P_n \sim (d-1)\eta^2$, where $d$ is number of the base states.

Note also, one should consider the possibility to increase the efficiency of the generation in terms of success probability by increasing the coupling parameter $\eta$, and as a consequence, the hybrid entanglement. It can be achieved by extending the time of interaction of the pump with the generated photons inside the crystal. Therefore, SPDC of the type-I matching in waveguide when all three waves propagate in the same direction is preferred. Another possibility to increase the coupling parameter $\eta$ can be associated with multiple interaction of the light waves with bulky crystal [33,34]. We can conjecture that each of the light waves passing through the crystal increases the coupling parameter by $\eta$ [34]. The coupling parameter may take value $N\eta$, where $N$ is the number of passes of light waves in the crystal. Another possibility to increase the hybrid entanglement may be related with use of optical parametric amplification (OPA). In and OPA the input, as rule, is two light beams. The OPA makes pumping weaker and amplifies other beams stronger and the interaction must be considered in depleted pump regime. It deserves separate investigation. We only note that it is sufficient to increase the coupling parameter $\eta$ holding the squeezing parameter $\alpha\eta \ll 1$ small enough to use the approximation (16a,16b). It is also worth noting, that, although probabilistic, the approach is not based on post-selection, both rather on heralding. This means the entangled qudit states remain fully available to further processing after their generation. Despite the small value of the coupling parameter $\eta$, the conditional generation can be implemented without photon number resolving detection.

**Appendix A. Matrix elements of the states in the pumping mode**

The displaced states are obtained by additional application of displacement operator $D(\alpha) = \exp(\alpha a^+ - \alpha^* a)$, where $\alpha$ is a displacement amplitude, to the input state [30,31]

$$|n,\alpha\rangle = D(\alpha)|n\rangle. \tag{A1}$$

The displaced number states (A1) are defined by two numbers: quantum discrete number $n$ and classical continuous parameter $\alpha$. Although, the number states and their displaced analogies have some quantum noise properties in common, they are not physically similar to each other. Additional classic parameter for the displaced states increases their energy by $|\alpha|^2$ $\left(n+|\alpha|^2\right)$ that allows us to consider the states as macroscopic in the case of $\alpha \gg 1$. Now, we are going to construct nonunitary infinite matrix $G$ connecting the unnormalized states $\left|\Phi_n^{(00)}\right\rangle$ (Eq. (10)) with displaced number states (A1)

$$\begin{Vmatrix} \left|\Phi_0^{(00)}\right\rangle \\ \left|\Phi_1^{(00)}\right\rangle \\ \left|\Phi_2^{(00)}\right\rangle \\ \cdots \\ \left|\Phi_n^{(00)}\right\rangle \\ \cdots \end{Vmatrix} = G \begin{Vmatrix} |0,\alpha\rangle \\ |1,\alpha\rangle \\ |2,\alpha\rangle \\ \cdots \\ |l,\alpha\rangle \\ \cdots \end{Vmatrix} = \begin{Vmatrix} g_{00} & g_{01} & g_{02} & \cdots & g_{0m} & \cdots \\ g_{10} & g_{11} & g_{12} & \cdots & g_{1m} & \cdots \\ g_{20} & g_{21} & g_{22} & \cdots & g_{2m} & \cdots \\ \cdots & \cdots & \cdots & \cdots & \cdots & \cdots \\ g_{n0} & g_{n1} & g_{n2} & \cdots & g_{nm} & \cdots \\ \cdots & \cdots & \cdots & \cdots & \cdots & \cdots \end{Vmatrix} \begin{Vmatrix} |0,\alpha\rangle \\ |1,\alpha\rangle \\ |2,\alpha\rangle \\ \cdots \\ |l,\alpha\rangle \\ \cdots \end{Vmatrix}, \tag{A2}$$

where

$$G = \begin{vmatrix} g_{00} & g_{01} & g_{02} & \cdots & g_{0m} & \cdots \\ g_{10} & g_{11} & g_{12} & \cdots & g_{1m} & \cdots \\ g_{20} & g_{21} & g_{22} & \cdots & g_{2m} & \cdots \\ \cdots & \cdots & \cdots & \cdots & \cdots & \cdots \\ g_{n0} & g_{n1} & g_{n2} & \cdots & g_{nm} & \cdots \\ \cdots & \cdots & \cdots & \cdots & \cdots & \cdots \end{vmatrix}. \tag{A3}$$

Here, the matrix $G$ is multiplied by infinite column vector of the displaced number states. So, first three states in the pumping mode often used here are written as

$$\left|\Phi_0^{(00)}\right\rangle = \sum_{m=0}^{\infty} g_{0m} |m,\alpha\rangle, \tag{A4}$$

$$\left|\Phi_1^{(00)}\right\rangle = \sum_{m=0}^{\infty} g_{1m} |m,\alpha\rangle, \tag{A5}$$

$$\left|\Phi_2^{(00)}\right\rangle = \sum_{m=0}^{\infty} g_{2m} |m,\alpha\rangle. \tag{A6}$$

So, real amplitudes $f_{2l,k}^{(00)}$ in equation (6) can be obtained from solution of sets of equations (7). The sets of equations (7) are not solved in exact form. Nevertheless, the solutions can be decomposed in series on small parameter $\eta \ll 1$. Substituting the analytical expressions for the amplitudes $f_{2l,k}^{(00)}$ recorded in the form of an expansion on the parameter $\eta$ in formulas (6) and using identities like

$$\alpha a^+ |0,\alpha\rangle = \exp\left(-|\alpha|^2/2\right) \sum_{l=1}^{\infty} \frac{l\alpha^l}{\sqrt{l!}} |l\rangle = |\alpha|^2 |0,\alpha\rangle + \alpha |1,\alpha\rangle, \tag{A7}$$

$$\alpha^2 a^{+2}|0,\alpha\rangle = \exp(-|\alpha|^2/2)\sum_{l=2}^{\infty}\frac{l(l-1)\alpha^l}{\sqrt{l!}}|l\rangle = |\alpha|^4|0,\alpha\rangle + 2\alpha|\alpha|^2|1,\alpha\rangle + \sqrt{2}\alpha^2|2,\alpha\rangle, \quad (A8)$$

$$\alpha^3 a^{+3}|0,\alpha\rangle = \exp(-|\alpha|^2/2)\sum_{l=3}^{\infty}\frac{l(l-1)(l-2)\alpha^l}{\sqrt{l!}}|l\rangle = |\alpha|^6|0,\alpha\rangle + 3\alpha|\alpha|^4|1,\alpha\rangle +$$
$$3\sqrt{2}\alpha^2|\alpha|^2|2,\alpha\rangle + \sqrt{3!}\alpha^3|3,\alpha\rangle \quad (A9)$$

one obtains the matrix elements $g_{nm}$ of the matrix $G$. We have

$$g_{00} = 1 - \frac{|\alpha|^2\eta^2}{2!} + \frac{5|\alpha|^4\eta^4 + |\alpha|^2\eta^4}{4!} - \frac{61|\alpha|^6\eta^6 + 35|\alpha|^4\eta^6 + |\alpha|^2\eta^6}{6!} + ..., \quad (A10)$$

$$g_{01} = \eta\left(-\frac{\alpha\eta}{2!} + \frac{10\alpha|\alpha|^2\eta^3}{4!} - \frac{183\alpha|\alpha|^4\eta^5 + 70\alpha|\alpha|^2\eta^5 + \alpha\eta^5}{6!} + ...\right), \quad (A11)$$

$$g_{02} = \eta^2\left(\frac{5\sqrt{2}\alpha^2\eta^2}{4!} - \frac{183\sqrt{2}\alpha^2|\alpha|^2\eta^4 + 35\sqrt{2}\alpha^2\eta^4}{6!} + ...\right), \quad (A12)$$

$$g_{03} = \eta^3\left(-\frac{61\sqrt{6}\alpha^3\eta^3}{6!} + ...\right), \quad (A13)$$

for zeroth vector row of the matrix (A3) or for the state (A4)

$$g_{10} = 1 - \frac{5|\alpha|^2\eta^2 + \eta^2}{3!} + \frac{61|\alpha|^4\eta^4 + 35|\alpha|^2\eta^4 + \eta^4}{5!} + ..., \quad (A14)$$

$$g_{11} = \eta\left(-\frac{5\alpha\eta}{3!} + \frac{122\alpha|\alpha|^2\eta^3 + 35\alpha\eta^3}{5!} + ...\right), \quad (A15)$$

$$g_{12} = \eta^2\left(-\frac{61\sqrt{2}\alpha^2\eta^2}{5!} + ...\right), \quad (A16)$$

for first vector row of the matrix (A3) or for the state (A5)

$$g_{20} = 1 - \frac{7|\alpha|^2\eta^2 + 3\eta^2}{3!} + \frac{331\alpha^2|\alpha|^4\eta^6 + 337\alpha^2|\alpha|^2\eta^6 + 36\alpha^2\eta^6}{3\times 5!} + ... \quad (A17)$$

$$g_{21} = \eta\left(-\frac{7\alpha\eta}{3!} + \frac{662\alpha^3|\alpha|^2\eta^5 + 337\alpha^3\eta^5}{3\times 5!} + ...\right), \quad (A18)$$

$$g_{22} = \eta^2\left(\frac{331\sqrt{2}\alpha^4\eta^4}{3\times 5!} + ...\right). \quad (A19)$$

for second vector row of the matrix (A2) or for the state (A6).

The matrix elements of tensor product of two states (17) are produced from the matrix elements $g_{nm}$ of the matrix $G$ and they are given by

$$p_{00}^{(2)} = g_{00}^2, \quad (A20)$$
$$p_{01}^{(2)} = \sqrt{2}g_{00}g_{01}, \quad (A21)$$
$$p_{02}^{(2)} = \sqrt{2}g_{00}g_{02}, \quad (A22)$$
$$p_{03}^{(2)} = g_{01}^2, \quad (A23)$$

for zeroth block

$$p_{10}^{(2)} = \sqrt{2}g_{00}g_{10}, \quad (A24)$$

$$p^{(2)}_{11} = g_{10}g_{01} + g_{00}g_{11}, \tag{A25}$$

$$p^{(2)}_{12} = g_{10}g_{01} - g_{00}g_{11}, \tag{A26}$$

$$p^{(2)}_{13} = g_{10}g_{02} + g_{00}g_{12}, \tag{A27}$$

$$p^{(2)}_{14} = g_{10}g_{02} - g_{00}g_{12}, \tag{A28}$$

$$p^{(2)}_{15} = \sqrt{2}g_{01}g_{11}, \tag{A29}$$

first block

$$p^{(2)}_{20} = \sqrt{2}g_{00}g_{20}, \tag{A30}$$

$$p^{(2)}_{21} = g_{20}g_{01} + g_{00}g_{21}, \tag{A31}$$

$$p^{(2)}_{22} = g_{20}g_{01} - g_{00}g_{21}, \tag{A32}$$

$$p^{(2)}_{23} = g_{20}g_{02} + g_{00}g_{22}, \tag{A33}$$

$$p^{(2)}_{24} = g_{20}g_{02} - g_{00}g_{22}, \tag{A34}$$

$$p^{(2)}_{25} = \sqrt{2}g_{01}g_{21}, \tag{A35}$$

$$p^{(2)}_{26} = g_{10}^2, \tag{A36}$$

$$p^{(2)}_{27} = \sqrt{2}g_{10}g_{11}, \tag{A37}$$

$$p^{(2)}_{28} = \sqrt{2}g_{10}g_{12}, \tag{A38}$$

$$p^{(2)}_{29} = g_{11}^2, \tag{A39}$$

and second blocks.

The same matrix elements $g_{nm}$ are converted to amplitudes $p^{(3)}_{nm}$ for the state (22) as

$$p^{(3)}_{00} = g_{00}^3, \tag{A40}$$

$$p^{(3)}_{01} = \sqrt{3}g_{00}^2g_{01}, \tag{A41}$$

$$p^{(3)}_{02} = \sqrt{3}g_{00}^2g_{02}, \tag{A42}$$

$$p^{(3)}_{03} = \sqrt{3}g_{00}g_{01}^2, \tag{A43}$$

for zeroth block

$$p^{(3)}_{10} = \sqrt{3}g_{00}^2g_{10}, \tag{A44}$$

$$p^{(3)}_{11} = g_{00}^2g_{11} + 2g_{00}g_{01}g_{10}, \tag{A44}$$

$$p^{(3)}_{12} = g_{00}^2g_{11} + \exp(-i\varphi)g_{00}g_{01}g_{10} + \exp(-i2\varphi)g_{00}g_{01}g_{10}, \tag{A46}$$

$$p^{(3)}_{13} = g_{00}^2g_{12} + 2g_{00}g_{10}g_{02}, \tag{A47}$$

$$p^{(3)}_{14} = g_{00}^2g_{12} + \exp(-i\varphi)g_{00}g_{10}g_{02} + \exp(-i2\varphi)g_{00}g_{10}g_{02}, \tag{A48}$$

$$p^{(3)}_{15} = g_{01}^2g_{10} + 2g_{00}g_{01}g_{11}, \tag{A49}$$

$$p^{(3)}_{16} = g_{00}g_{01}g_{11} + \exp(-i\varphi)g_{00}g_{01}g_{11} + \exp(-i2\varphi)g_{01}^2g_{10}, \tag{A50}$$

$$p^{(3)}_{17} = g_{00}g_{01}g_{11} + \exp(-i\varphi)g_{01}^2g_{10} + \exp(-i2\varphi)g_{00}g_{01}g_{11}, \tag{A51}$$

first block

$$p^{(3)}_{20} = \sqrt{3}g_{00}^2g_{20}, \tag{A52}$$

$$p^{(3)}_{21} = g_{00}^2g_{21} + 2g_{00}g_{01}g_{20}, \tag{A53}$$

$$p^{(3)}_{22} = g_{00}^2g_{21} + \exp(-i\varphi)g_{00}g_{01}g_{20} + \exp(-i2\varphi)g_{00}g_{01}g_{20}, \tag{A54}$$

$$p^{(3)}_{23} = g_{00}^2g_{22} + 2g_{00}g_{02}g_{20}, \tag{A55}$$

$$p^{(3)}_{24} = g_{00}^2g_{22} + \exp(-i\varphi)g_{00}g_{02}g_{20} + \exp(-i2\varphi)g_{00}g_{02}g_{20}, \tag{A56}$$

$$p_{25}^{(3)} = g_{01}^2 g_{20} + 2g_{00}g_{01}g_{21}, \tag{A57}$$

$$p_{26}^{(3)} = g_{00}g_{01}g_{21} + \exp(-i\varphi)g_{00}g_{01}g_{21} + \exp(-i2\varphi)g_{01}^2 g_{20}, \tag{A58}$$

$$p_{27}^{(3)} = g_{00}g_{01}g_{21} + \exp(-i\varphi)g_{01}^2 g_{20} + \exp(-i2\varphi)g_{00}g_{01}g_{21}, \tag{A59}$$

$$p_{28}^{(3)} = \sqrt{3}g_{00}g_{10}^2, \tag{A60}$$

$$p_{29}^{(3)} = g_{10}^2 g_{01} + 2g_{00}g_{10}g_{11}, \tag{A61}$$

$$p_{210}^{(3)} = g_{00}g_{10}g_{11} + \exp(-i\varphi)g_{10}^2 g_{01} + \exp(-i2\varphi)g_{00}g_{10}g_{11}, \tag{A62}$$

$$p_{211}^{(3)} = g_{00}g_{10}g_{11} + \exp(-i\varphi)g_{00}g_{10}g_{11} + \exp(-i2\varphi)g_{10}^2 g_{01}, \tag{A63}$$

$$p_{212}^{(3)} = g_{10}^2 g_{02} + 2g_{00}g_{10}g_{12}, \tag{A64}$$

$$p_{213}^{(3)} = g_{00}g_{10}g_{12} + \exp(-i\varphi)g_{10}^2 g_{02} + \exp(-i2\varphi)g_{00}g_{10}g_{12}, \tag{A65}$$

$$p_{214}^{(3)} = g_{00}g_{10}g_{12} + \exp(-i\varphi)g_{00}g_{10}g_{12} + \exp(-i2\varphi)g_{10}^2 g_{02}, \tag{A66}$$

$$p_{215}^{(3)} = g_{00}g_{11}^2 + 2g_{01}g_{10}g_{11}, \tag{A67}$$

$$p_{216}^{(3)} = g_{00}g_{11}^2 + \exp(-i\varphi)g_{01}g_{10}g_{11} + \exp(-i2\varphi)g_{01}g_{10}g_{11}, \tag{A68}$$

and second blocks.

**Appendix B. Transformations in pumping modes**

Consider unitary transformations of the states in pumping mode. The beam splitter (31) transforms the input states (18,20) to

$$BS\big(|0,\alpha\rangle_{p_1}|0,\alpha\rangle_{p_2}\big) = |0\rangle_{p_1}|0,\sqrt{2}\alpha\rangle_{p_2}, \tag{B1}$$

$$BS\big(|1,\alpha\rangle_{p_1}|1,\alpha\rangle_{p_2}\big) = \big(-|2\rangle_{p_1}|0,\sqrt{2}\alpha\rangle_{p_2} + |0\rangle_{p_1}|2,\sqrt{2}\alpha\rangle_{p_2}\big)\big/\sqrt{2}, \tag{B2}$$

$$BS|\varphi_+\rangle_{p_1 p_2} = |0\rangle_{p_1}|1,\sqrt{2}\alpha\rangle_{p_2}, \tag{B3}$$

$$BS|\varphi_-\rangle_{p_1 p_2} = -|1\rangle_{p_1}|0,\sqrt{2}\alpha\rangle_{p_2}, \tag{B4}$$

$$BS|\sigma_+\rangle_{p_1 p_2} = \big(|2\rangle_{p_1}|0,\sqrt{2}\alpha\rangle_{p_2} + |0\rangle_{p_1}|2,\sqrt{2}\alpha\rangle_{p_2}\big)\big/\sqrt{2}, \tag{B5}$$

$$BS|\sigma_-\rangle_{p_1 p_2} = -|1\rangle_{p_1}|1,\sqrt{2}\alpha\rangle_{p_2}. \tag{B6}$$

Action of the unitary transformation (37) on the states (27-29) leads to

$$U\big(|0,\alpha\rangle_{p_1}|0,\alpha\rangle_{p_2}|0,\alpha\rangle_{p_3}\big) = |0,\sqrt{3}\alpha\rangle_{p_1}|0\rangle_{p_2}|0\rangle_{p_3}, \tag{B7}$$

$$U|\varphi_0\rangle_{p_1 p_2 p_3} = |1,\sqrt{3}\alpha\rangle_{p_1}|0\rangle_{p_2}|0\rangle_{p_3}, \tag{B8}$$

$$U|\varphi_{-\phi}\rangle_{p_1 p_2 p_3} = |1,\sqrt{3}\alpha\rangle_{p_1}|1\rangle_{p_2}|0\rangle_{p_3}, \tag{B9}$$

$$U|\varphi_{-2\phi}\rangle_{p_1 p_2 p_3} = |1,\sqrt{3}\alpha\rangle_{p_1}|0\rangle_{p_2}|1\rangle_{p_3}, \tag{B10}$$

$$U|\sigma_0\rangle_{p_1 p_2 p_3} = \big(|2,\sqrt{3}\alpha\rangle_{p_1}|0\rangle_{p_2}|0\rangle_{p_3} + \sqrt{2}|0,\sqrt{3}\alpha\rangle_{p_1}|1\rangle_{p_2}|1\rangle_{p_3}\big)\big/\sqrt{3}, \tag{B11}$$

$$U|\sigma_{-\phi}\rangle_{p_1 p_2 p_3} = \big(|0,\sqrt{3}\alpha\rangle_{p_1}|0\rangle_{p_2}|2\rangle_{p_3} + \sqrt{2}|1,\sqrt{3}\alpha\rangle_{p_1}|1\rangle_{p_2}|0\rangle_{p_3}\big)\big/\sqrt{3}, \tag{B12}$$

$$U|\sigma_{-2\phi}\rangle_{p_1 p_2 p_3} = \big(|0,\sqrt{3}\alpha\rangle_{p_1}|2\rangle_{p_2}|0\rangle_{p_3} + \sqrt{2}|1,\sqrt{3}\alpha\rangle_{p_1}|0\rangle_{p_2}|1\rangle_{p_3}\big)\big/\sqrt{3}, \tag{B13}$$

$$U|\zeta_0\rangle_{p_1p_2p_3} = \left(\sqrt{2}|2,\sqrt{3}\alpha\rangle_{p_1}|0\rangle_{p_2}|0\rangle_{p_3} - |0,\sqrt{3}\alpha\rangle_{p_1}|1\rangle_{p_2}|1\rangle_{p_3}\right)/\sqrt{3}, \tag{B14}$$

$$U|\zeta_{-\phi}\rangle_{p_1p_2p_3} = \exp(i2\phi)\left(\sqrt{2}|0,\sqrt{3}\alpha\rangle_{p_1}|0\rangle_{p_2}|2\rangle_{p_3} - |1,\sqrt{3}\alpha\rangle_{p_1}|1\rangle_{p_2}|0\rangle_{p_3}\right)/\sqrt{3}, \tag{B15}$$

$$U|\zeta_{-2\phi}\rangle_{p_1p_2p_3} = \exp(i\phi)\left(\sqrt{2}|0,\sqrt{3}\alpha\rangle_{p_1}|2\rangle_{p_2}|0\rangle_{p_3} - |1,\sqrt{3}\alpha\rangle_{p_1}|0\rangle_{p_2}|1\rangle_{p_3}\right)/\sqrt{3}. \tag{B16}$$

**Acknowledgement**

The work was supported by Act 211 Government of the Russian Federation, contract № 02.A03.21.0011.

**List of figures**

**Figure 1(a,b)**
Schematic representation of hybrid entanglement test and conditional generation of maximally entangled states of two (a) qubits and (b) qutris. $U(\varphi)$ means discrete Fourier transform (37).

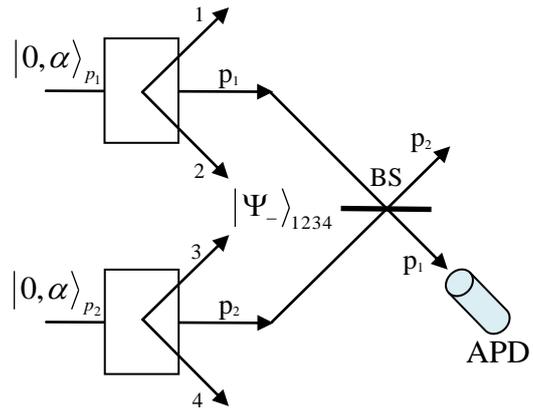

(a)

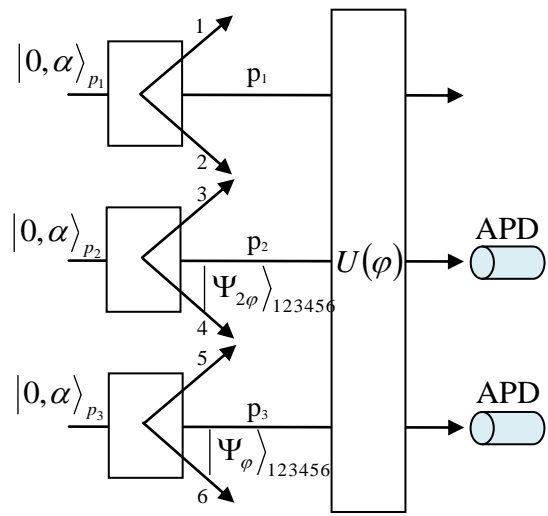

(b)

**Figure 1(a,b)**